\documentclass[%
 reprint,
 amsmath,amssymb,
 aps,
 prl
]{revtex4-2}

\usepackage{graphicx}
\usepackage{dcolumn}
\usepackage{bm}

\usepackage{tabularx}

\usepackage[deletedmarkup=xout]{changes}

\definechangesauthor[color=violet]{DNS}

\definechangesauthor[color=red]{Péter}

\definechangesauthor[color=blue]{Gábor}

\begin{document}

\preprint{APS/123-QED}

\title{Dynamic Length Scale and Weakest Link Behavior in Crystal Plasticity \footnote{See published version at: https://journals.aps.org/prmaterials/abstract/10.1103/PhysRevMaterials.7.013604}}

\author{D\'enes Berta}
\author{G\'abor P\'eterffy}
\author{P\'eter Dus\'an Isp\'anovity}
\affiliation{E\"otv\"os Lor\'and University, Department of Materials Physics, P\'azm\'any P\'eter s\'et\'any 1/A, 1117 Budapest, Hungary}


\begin{abstract}

Irreversible deformation of crystals is often characterized by stochastic scale-free distributed intermittent local plastic bursts. Quenched obstacles with short-range interaction were found to limit the size of these events, that was termed as transition from wild to mild fluctuations. Here we show by analysing the local yield thresholds in a discrete dislocation model that a dynamic length scale can be introduced based on weakest link principles and this scale characterizes the extension of plastic events. The interplay between long-range dislocation interactions and short-range quenched disorder is found to destroy scale-free dynamical correlations, thus, leading to event localization (that is, shortening of the length scale) which explains the crossover between the wild and mild regimes. Several methods are presented to determine the dynamic length scale which can be generalized to other types of heterogeneous materials.

\end{abstract}

\maketitle

\section{Introduction}

Plastic behavior at micron and sub-micron scales differs profoundly from that of bulk materials: Significant size-related hardening \cite{fleck1994strain, uchic2004sample, uchic2009plasticity} and intermittent stochastic strain bursts \cite{dimiduk2006scale, csikor2007dislocation} can be observed. The latter causes unpredictable plasticity and staircase-like patterns in the stress-strain curves in contrast to the smooth curves of bulk specimens. In crystalline materials plastic events are avalanche-like rearrangements of topological crystallographic defects (dislocations). These events are also present in bulk materials, as shown by studies investigating acoustic signals emitted by these avalanches \cite{weiss1997acoustic, miguel2001intermittent, weiss2003three, weiss2015mild, ispanovity2022dislocation}. In amorphous solids and foams deformation is characterized by similar fluctuations but the irreversible units of plasticity are shear transformation zones \cite{spaepen1977microscopic, falk1998dynamics}, and T1 events \cite{kabla2003local, dennin2004statistics, tainio2021predicting}, respectively. Thus, one may conclude that all these heterogeneous materials exhibit substantially analogous, stochastic plastic response.

Indeed, some authors have advanced the idea that plasticity exhibits universality in a wide range of materials and scales up to that of earthquakes \cite{kagan2010earthquake, uhl2015universal, dahmen2017mean}. In crystalline solids the picture is more complex since microstructure has a crucial impact on the critical behavior. On the one hand, materials with HCP structure, where practically single slip deformation takes place, exhibit large, scale-free fluctuations \cite{miguel2001intermittent, weiss2015mild, ispanovity2022dislocation}. On the other hand, when the dynamics of dislocations gets more complex, e.g., at multiple slip in FCC or BCC structures or by the addition of solute atoms that hinder the motion of dislocations with short-range forces, fluctuations may get bounded or may even disappear (however, we do note that long-tailed criticality is not always destroyed by defects \cite{rizzardi2022intermittent}). This phenomenon, observed with the help of acoustic emission as well as micropillar compression experiments, was termed ``wild to mild'' transition  \cite{weiss2015mild, zhang2017taming, weiss.2021r}. Clearly, the situation is even more complex as it is also affected by specimen size as ``smaller is wider'', and it was found to be the result of the competition of an external length scale (due to the finite specimen size) and some internal length scale (due to microstructural disorder) \cite{zhang2017taming, weiss.2021r}. It has been shown recently that further enrichment of the phenomenon is caused by additional mechanisms, such as the effect of grain structure or the Portevin–Le Chatelier effect \cite{lebedkina2018correlation}. Analogous conclusions were drawn from simulation of the dynamics of straight edge dislocation ensembles in single slip. With the absence of quenched disorder the system exhibits criticality even at zero applied stress \cite{ispanovity2014avalanches,lehtinen2016glassy}, however, the inclusion of point defects with short-range interaction leads to a subcritical state with bounded avalanches at small applied stresses and changes the universality class of the yielding transition \cite{ovaska2015quenched, salmenjoki2020plastic}. Although a lot of modelling activities, involving dislocation dynamics simulations, cellular automaton plasticity simulations as well as stochastic crystal plasticity simulations were devoted to the issue of dislocation avalanches and the corresponding universality classes (see, e.g., \cite{cui2017influence, song2019universality, zhang2020variety, yu2021stochastic}) the precise definition of the length-scale that controls fluctuations remains elusive. In this paper, therefore, we intend to analyse the aforementioned wild to mild transition on the model system of edge dislocations and aim at providing a proper definition of the dynamic length-scale that controls fluctuations and linking this scale to microstructural features and understanding its role in the localization of plastic slip. The focus will be on the microplastic regime, that is, plasticity taking place at small loads below the yield stress, so, investigating the critical behavior associated with the yielding transition is out of the scope of the present paper.

Microplasticity is often explained based on weakest-link theory both for crystalline \cite{parthasarathy2007contribution, el2009role, ispanovity2013average, derlet2015probabilistic, derlet2016stress} and amorphous matter \cite{lai2008bulk, ye2010extraction}. The general assumption is that as load increases the weakest spots of the material get subsequently activated. It can be assumed that microstructural heterogeneity affects plasticity through the variations in local strength, and this idea led to the development of mesoscopic elasto-plastic models for both amorphous \cite{baret2002extremal, talamali2011avalanches, liu2016driving, budrikis2017universal, nicolas2017deformation, tyukodi2019avalanches, popovic2021thermally, khirallah2021yielding} and crystalline \cite{zaiser2005fluctuation, papanikolaou2012quasi} materials. In these models whenever the local stress at a given cell exceeds the local threshold, plastic strain is accumulated giving rise to the anisotropic redistribution of the internal stress, which may lead to subsequent activation of another cell. These general models can, among others, account for the avalanche dynamics characteristic of heterogeneous materials.

The above mentioned weakest-link argument is straightforward if plasticity is local, however, its possible non-locality was pointed out both for crystalline \cite{ispanovity2014avalanches, ispanovity2017role} and amorphous solids \cite{lerner2009locality}. Thus, a fundamental, so far not addressed, question is how to select the size of the sub-volume [the representative volume element (RVE)] which is represented by a \emph{local yield stress} value (i.e., the stress threshold of plastic yielding). In this paper we address precisely this issue, that is, the dependence of the local yield stress statistics on the size of the local sub-volume using a general model for crystalline plasticity (general here refers to the fact that this model focuses on the most general properties of dislocation dynamics such as long-range mutual interactions and dissipative motion of dislocations and specific properties dependent on the crystal structure or temperature, such as cross-slip or core effects, are not considered). As it will be shown, the analysis will allow us to identify the corresponding dynamic length-scale discussed above and to test whether and how the weakest-link picture is realized. To this end, we will study the statistical properties of the local yield stress, since it has been shown to have a profound connection to the loci of plastic events during global loading of model amorphous solids \cite{patinet2016connecting, barbot2018local} and has also been adapted for crystalline materials \cite{ovaska2017excitation}.

\section{Numerical model of dislocation dynamics}

To investigate the problem at hand a two-dimensional (2D) discrete dislocation dynamics (DDD) model is used. The system consists of $N=1024$ edge dislocations that are straight, parallel, and lie on parallel slip planes. The positions $\bm r_i$ of the dislocations are tracked on the $xy$ plane perpendicular to the dislocation lines. Let the Burgers vectors be parallel with the axis $x$: $\bm b_i = (\pm b, 0)$ with the same number of types + and --. The simulation cell is square-shaped with periodic boundary conditions \cite{kuykendall2013conditional, peterffy2020efficient} and contains varying number $N_\mathrm{p}$ of immobile point defects. Let $Q$ denote the ratio of these constituents: $Q = N_\mathrm{p}/N$. Here $0\leq Q\leq 10$. The motion of dislocations is determined by the forces acting on them caused by long-range dislocation-dislocation and short-range dislocation-point defect interactions and an empirical mobility law \cite{anderson2017theory,supplemental}. This model focuses primarily on the effect of the long-range elastic interaction between dislocations and its interplay with the quenched disorder and the related physics. We emphasize that the model certainly cannot account for several 3D dislocation mechanisms, such as dislocation source truncation or starvation that may play an important role at small specimen sizes. It is also mentioned that other models have also been used to model 2D crystal plasticity, such as one based on Landau theory \cite{baggio2019landau}.

In this paper stresses will be measured in units of $\tau_0 = \frac{\mu b\sqrt{\rho}}{2\pi(1-\nu)}$, the interaction stress between two dislocations at a distance of the average dislocation spacing. Here $\mu$, $\rho=N/L^2$ and $\nu$ are the shear modulus, the dislocation density and the Poisson's ratio, respectively. Initial configurations were obtained by letting systems of randomly positioned dislocations (sampled from 2D uniform distribution) relax at zero applied stress. (We note that choosing the initial configuration according to a restricted random configuration proposed by Wilkens \cite{wilkens1970fundamental} does not affect the results, for details  see \cite{supplemental}.) To determine the local yield stresses, subsystems were then locally loaded with a slowly increasing homogeneous external stress acting on dislocations within the box, while the outer ones were kept fixed. The plastic event is considered to set on if any individual dislocation exceeds a certain velocity threshold \cite{supplemental}. The 2D DDD is a strongly simplified model of crystal plasticity, so, it is not meant to reproduce precise values of stresses measured in experiments for real materials, however, it may still be of interest to compare these values. One way to test that is to compare the flow stresses of the same 2D DDD systems obtained earlier \cite{szabo2015plastic} with experimental values of single crystals. On the experimental side, the yield stress is expressed by the Taylor-relation $\tau_\text{y}=\alpha\mu b\sqrt{\rho}$, with the dimensionless parameter $\alpha$ found to be around $0.1-0.4$ for single crystals \cite{mughrabi2016alpha}. In the 2D DDD systems a flow stress of $(0.9 \pm 0.3)\tau_0$ was obtained (see Figs.~4 and 5 in \cite{szabo2015plastic}). Assuming $\nu = 0.35$ the $\alpha$ parameter from the simulations is $0.22 \pm 0.07$. This means the values are, in fact, in surprisingly good accordance. However, we stress again, that providing exact yield stress values is not expected from this toy model.

\section{Loading protocol to determine local yield stresses}

\begin{figure*}[!ht]
    \begin{center}
    \includegraphics[width=\textwidth]{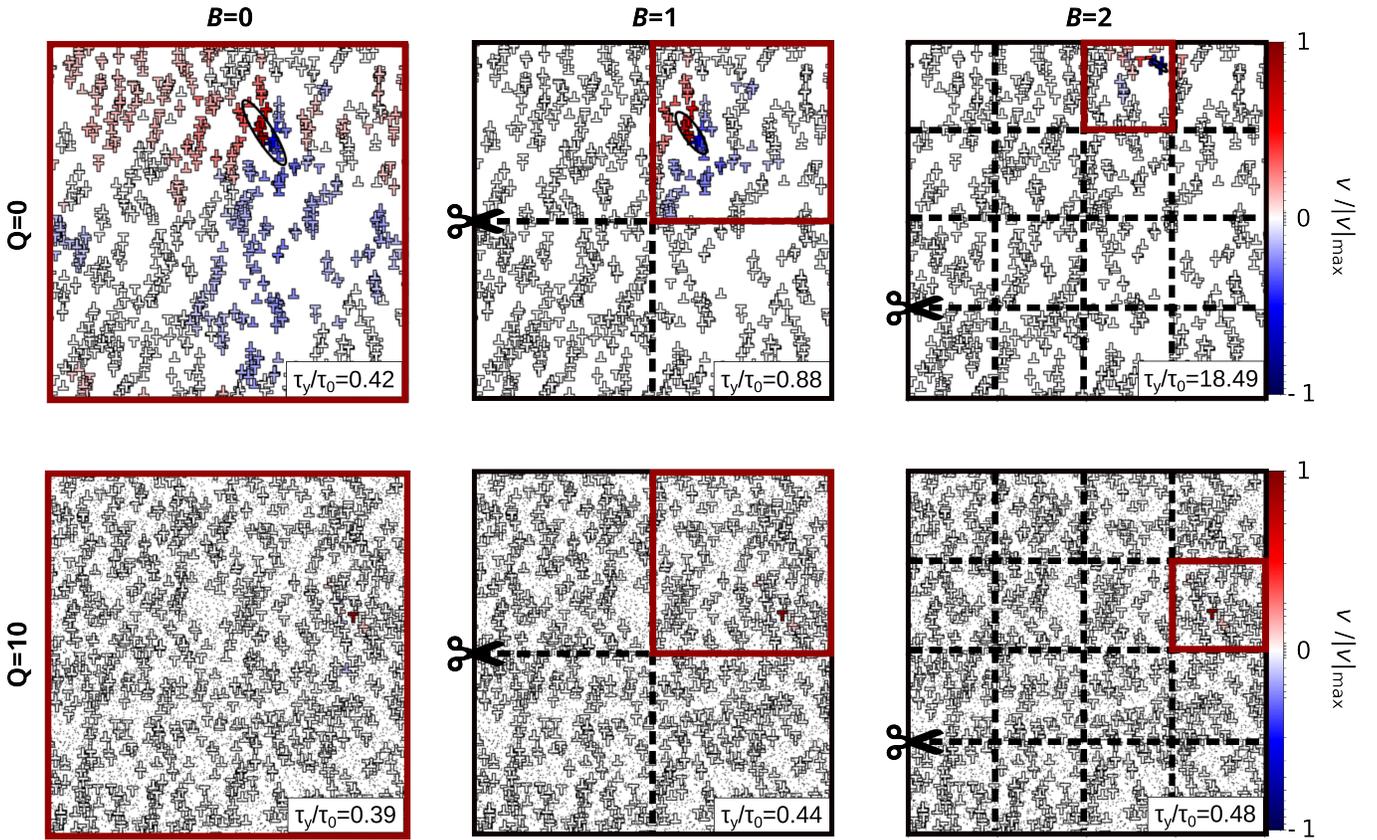}
    \end{center}
    \caption{\label{fig:box_division} The initial state of the first avalanche under global ($B=0$) and local ($B=1,2$) loading of the red box in a pure ($Q=0$) dislocation system (upper row) and a system rich in point defects ($Q=10$, bottom row). Subsystems (boxes) are obtained by recursive division of the simulation cell. The active dislocations are colored according to the magnitude and direction of the velocity $v$: red-colored dislocations are moving to the right, blue ones are moving to the left and the white ones are almost still. Note the similarities in the dislocation velocities and the local yield stresses between boxes at levels $B=0$ and $B=1$ in the case of $Q=0$. Local yield stresses are even more similar in the case of $Q=10$ where the localized event is not affected much by the box division.}
\end{figure*}

In previous works focusing on amorphous solids, spherical regions were loaded \cite{patinet2016connecting, barbot2018local, ruan2022predicting}. These spheres were centered on atoms and may overlap. Another method used for crystalline solids is loading dislocations individually \cite{ovaska2017excitation}. In our simulations the subsystems are chosen differently: square grids of different resolutions are created and the (disjoint) grid cells are loaded separately, that is, external load is only applied to dislocations that are within the given cell and the other dislocations are kept fixed (see the two representative systems of $Q=0$ and $Q=10$ in Fig.~\ref{fig:box_division}). This external stress applied is the same for all dislocations within the box and it is increased quasi-statically until the onset of the first avalanche. Three factors led to this choice: Firstly, equilibrium dislocation densities are much more heterogeneous than atomic density in amorphous solids, thus, locating the centers on dislocations (or selecting individual dislocations) necessarily weights the local yield threshold statistics quite unevenly. Secondly, this selection is also motivated by nanoindentation experiments commonly used for measuring local hardness. Here a local volume is loaded (although unevenly), thus, loading of a finite local volume (being of spherical, rectangular or any other simple shape) seems a more natural choice then exciting individual dislocations. (Note, that here we do not intend to model nanoindentation, this experimental technique merely serves as a motivation for our loading protocol in the simulations.) Thirdly, local yield stress is an important variable in mesoscale simulations \cite{baret2002extremal, talamali2011avalanches, liu2016driving, budrikis2017universal, popovic2021thermally} and in continuum dislocation field theories as well \cite{xia2015preliminary, ispanovity2017role, sudmanns2019dislocation, ispanovity2020emergence, wu2021cell, zoller2021microstructure} and the numerical solution of these models are performed on rectangular grids with lattice spacing (that is, spatial resolution) as parameter. Thus, local yield stress statistics (distribution, spatial correlations) computed on such grids could be directly applied as input for such continuum models. The grid is obtained starting from the whole simulation cell which is then cut in half recursively both vertically and horizontally.
The number of subsequent division steps is denoted by $B$ (see Fig.~\ref{fig:box_division}).
This procedure is continued until empty boxes (without dislocations) start to appear (after $B=3$ in our case) \cite{supplemental}.

\begin{figure}[!ht]
    \centering
    \includegraphics[width=\columnwidth]{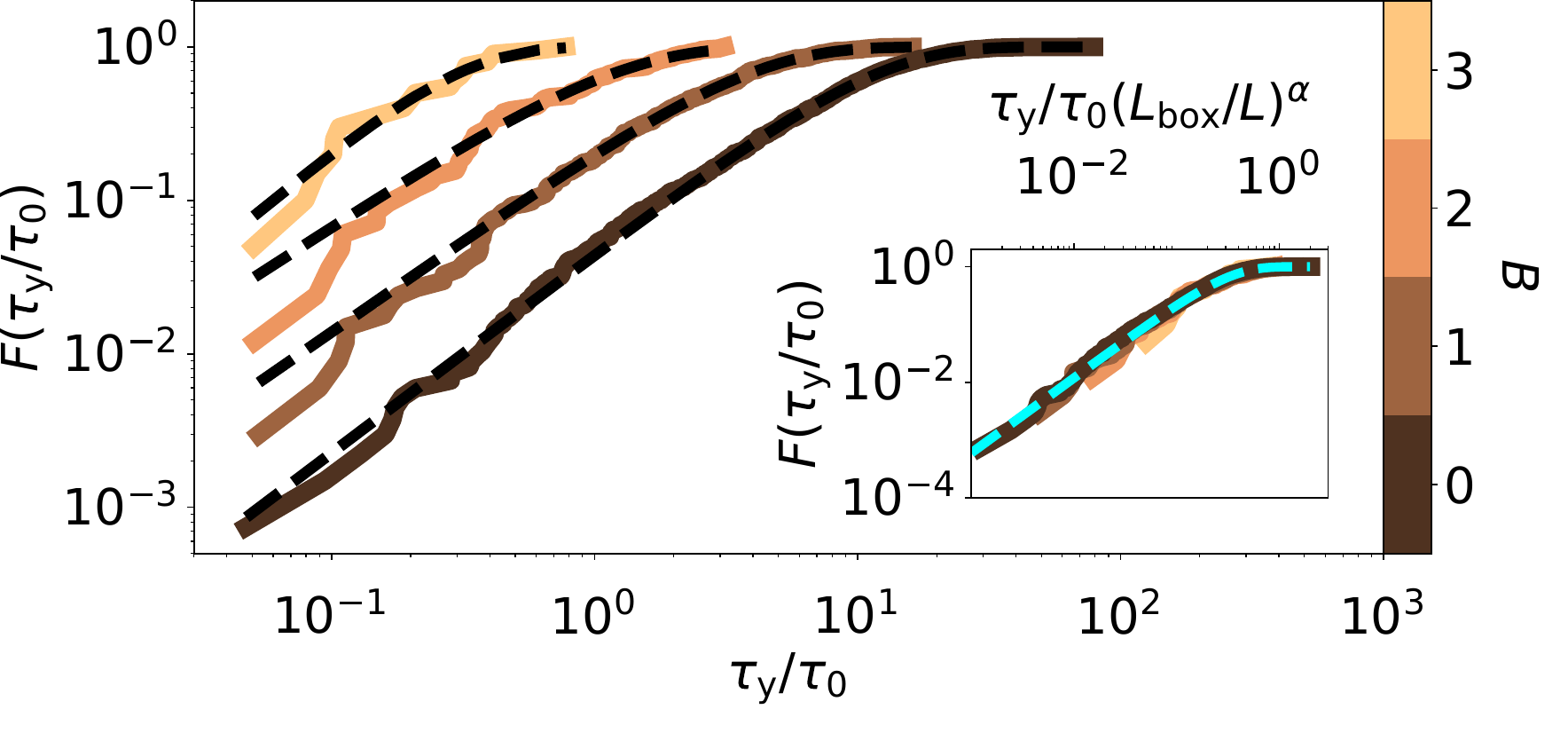}
    \caption{The CDF $F$  of the local yield stress $\tau_\text{y}$ at different $B$ values at $Q=10$. The inset shows scaling collapse and the fitted Weibull distribution (turquoise) according to Eqs.~(\ref{eq:weibull}) and (\ref{eq:lambda_scaling}). Collapse obtained with $\alpha = 1.8 \pm 0.1$.}
    \label{fig:distributions}
\end{figure}

\section{Local yield stress statistics}

Figure \ref{fig:distributions} shows the distributions of local yield stresses obtained for different sizes $B$ at $Q=10$ (for other values of $Q$ see \cite{supplemental}). Since the strength of the weakest-links is assumed to determine the yield threshold in boxes containing numerous links, one may expect to get an extremal probability distribution. In particular, if in the small strength limit $F_\text{link}(\tau_\text{y}) \propto \tau_\text{y}^k$, $F_\text{link}$ being the CDF of the yield threshold $\tau_\mathrm{y}$, then the emergent extremal probability distribution is of Weibull type \cite{weibull1939statistical, weibull1951wide, derlet2015probabilistic, derlet2015universal, ispanovity2017role} with a CDF
\begin{equation}
    F(\tau_\mathrm{y})=1-\mathrm{exp}\left[-\left(\frac{\tau_\mathrm{y}}{\lambda}\right)^k\right].
    \label{eq:weibull}
\end{equation}
Here $k$ and $\lambda$ are the so-called \emph{shape} and \emph{scale parameters}, respectively. As seen in Fig.~\ref{fig:distributions} these Weibull distributions are reproduced by our simulations (with a shape parameter tending from $k = 1.6\pm0.05$ at $Q=0$ to $k=1.3\pm0.03$ at $Q=10$). Additionally, the scaling collapse seen in the inset shows that the scale parameter (proportional to the average yield stress) scales with the linear subbox size $L_\mathrm{box} = 2^{-B} L$ with an exponent $\alpha$:
\begin{equation}
    \lambda \propto L_\mathrm{box}^{-\alpha}.
    \label{eq:lambda_scaling}
\end{equation}

\begin{figure}[!ht]
    \centering
    \includegraphics[width=\columnwidth]{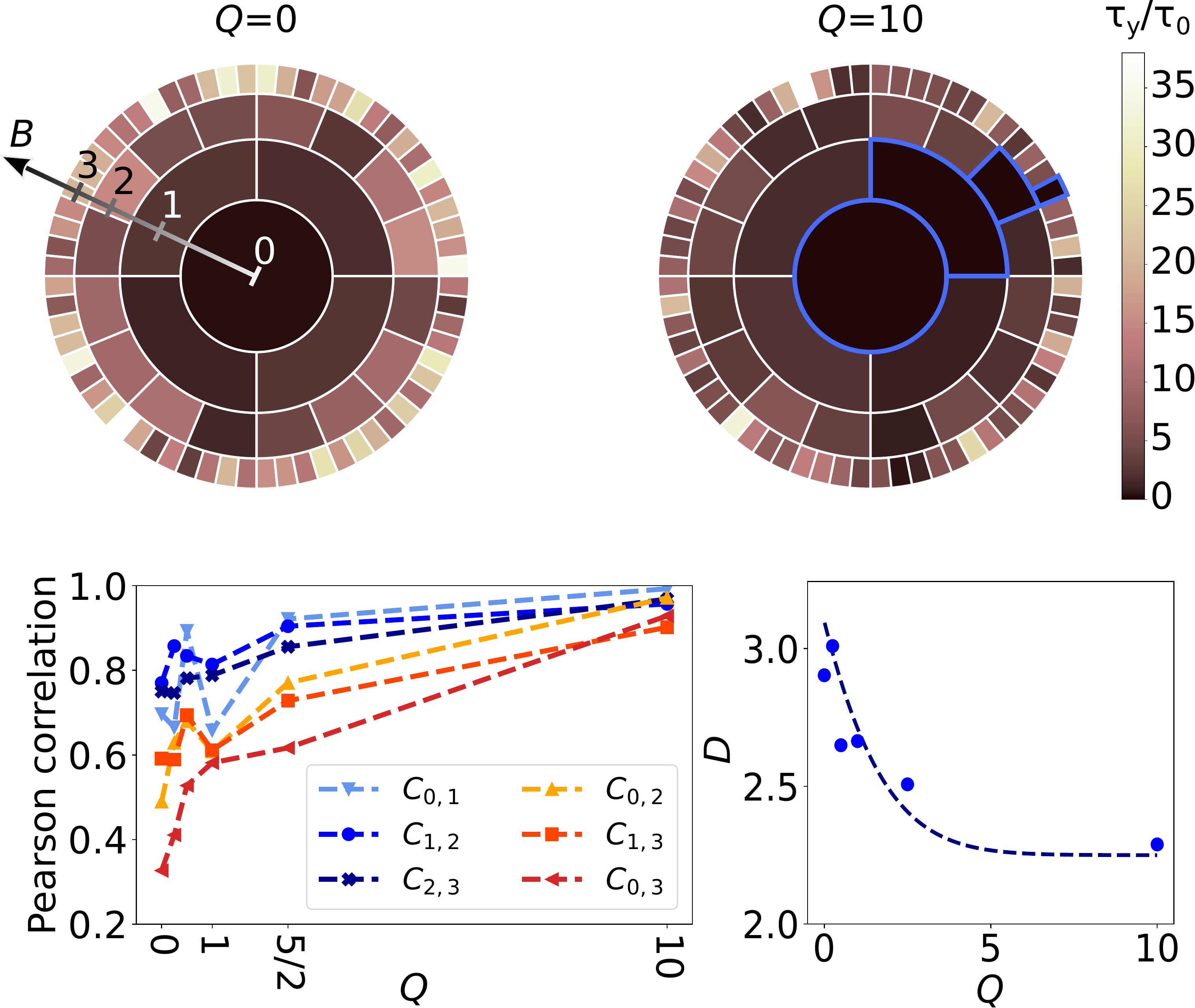}
    \begin{picture}(0,0)
    \put(-120,215){\sffamily{a)}}
    \put(10,215){\sffamily{b)}}
    \put(-120,110){\sffamily{c)}}
    \put(25,110){\sffamily{d)}}
    \end{picture}
    \caption{\label{fig:round_yield} Yield stresses of single systems at $Q=0$ (a) and $Q=10$ (b). One resolution $B$ corresponds to one ring in the pie chart. The box-subbox relations are represented by radial adjacency. The correlation of yield stresses of parent boxes and their softest subboxes is remarkably high, particularly in the $Q=10$ case. A sequence of the softest subboxes at different levels is highlighted with blue contour in panel b). (c): Pearson correlations $C_{m,n}$ defined by Eq.~(\ref{eq:correlation}) for different values of $Q$. (d): The link-dimension $D$ defined by Eq.~(\ref{eq: dimension def}) against point defect concentration $Q$. The dashed curve is just a guide to the eye.}
\end{figure}

If the weakest-link picture is realized (as assumed in the mesoscale plasticity models described above) the yield stress of each box is equal to that of its softest subbox. To test whether it is indeed the case here, box-subbox modified Pearson correlations are computed according to
\begin{equation}
    C_{m,n}=\frac{\left\langle \tau_{\text{y},m}^i \tau_{\text{y},m,n}^{i} \right\rangle - \left\langle \tau_{\text{y},m}^i \right\rangle \left\langle \tau_{\text{y},m,n}^{i} \right\rangle}{\sqrt{\left\langle (\tau_{\text{y},m}^i)^2 \right\rangle - \left\langle \tau_{\text{y},m}^i \right\rangle^2}\sqrt{\left\langle (\tau_{\text{y},m,n}^{i})^2 \right\rangle - \left\langle \tau_{\text{y},m,n}^{i} \right\rangle^2}},
    \label{eq:correlation}
\end{equation}
where $\tau_{\text{y},m}^i$ denotes the local yield threshold of parent box $i$ at level $B=m$ and $\tau_{\text{y},m,n}^{i}$ stands for the \emph{minimum} of the yield stresses of the subboxes at level $B=n>m$ of the parent box $i$. The angle brackets denote expected value over parent boxes $i$ averaged over all systems at given $Q$. There are indeed high correlations as seen in Fig.~\ref{fig:round_yield}(c). This is particularly true in systems rich in point defects with values larger than 0.9. At $Q=0$, however, the correlations are somewhat lower, especially in the case of distant levels $m$ and $n$.

Figures \ref{fig:round_yield}(a) and (b) also clearly show how systems with point defects outperform the pure dislocation systems. The yield stress maps of two representative systems with $Q=0$ and $Q=10$ are condensed into single pie charts. In the latter a very prominent chain of weakest links is highlighted with blue contour. In the pure system, however, the weakest-link behavior is not that apparent.

According to extreme value theory the scale parameter $\lambda$ is related to the number $N_\mathrm{link}$ of links as $\lambda \propto N_\mathrm{link}^{-1/k}$ \cite{ispanovity2017role}. This with Eq.~(\ref{eq:lambda_scaling}) yields
\begin{equation}
    N_\mathrm{link}(L_\mathrm{box})\propto L_\mathrm{box}^{D} = L_\mathrm{box}^{k\alpha},
    \label{eq: dimension def}
\end{equation}
where the \emph{link-dimension} $D$ was introduced as $D=k\alpha$. According to Fig.~\ref{fig:round_yield}(d) the systems show an anomalous, super-extensive scaling of the number of links with $D$ typically being between $2$ and $3$. The highest values of $D$ occur in systems at low $Q$ and as $Q$ increases $D$ tends to $2$ corresponding to extensive scaling. This together with the particularly high correlations suggests that the introduction of quenched disorder localizes the plastic events.

\section{Distribution of event sizes and the emergent length scale}

\begin{figure}[ht!]
    \centering
    \includegraphics[width=\columnwidth]{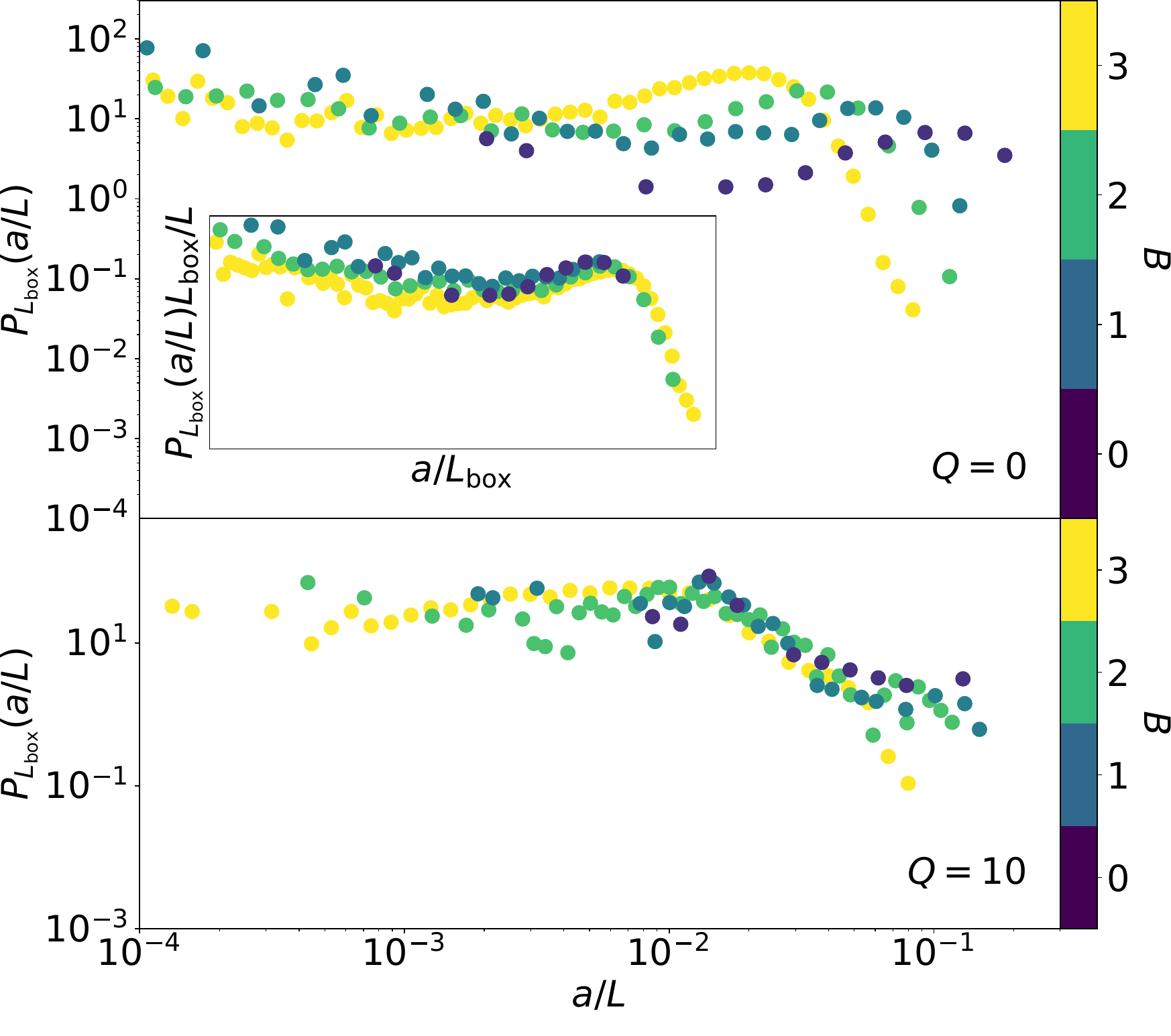}
    \caption{The PDF $P_{L_\mathrm{box}}$ of link size $a/L$ for pure ($Q=0$) and defect-rich ($Q=10$) systems. At $Q=0$ the characteristic link size varies with box size, while at $Q=10$ different-sized boxes behave similarly. The inset shows curve-collapse indicating the extensive scaling of link size at $Q=0$.}
    \label{fig:pdf_a}
\end{figure}

To quantify localization, the velocities of dislocations at the onset of the first plastic event were computed. As shown in the representative cases of Fig.~\ref{fig:box_division}, the most active dislocations are located in a finite region. The corresponding linear size $a$ was estimated by the semi-major axis of an ellipse fitted to the active region \cite{supplemental} (which is shown in the $Q=0$ cases of Fig. \ref{fig:box_division} where the ellipses are large enough to be visible). Figure \ref{fig:pdf_a} plots the distribution $P_{L_\mathrm{box}}({a/L})$ of the event size for different box sizes $L_\text{box}$ and concentration $Q$. As seen, for $Q=0$ the distribution strongly depends on the box size as size $a$ can always approach $L_\text{box}$. In addition, the distributions obey a simple scaling property
\begin{equation}
    P_{L_\text{box}}(a) = p(a/ L_\text{box}) / L_\text{box}, \qquad (Q=0)
\end{equation}
with a suitable function $p$. This suggests that in the $Q=0$ limit there is no length-scale associated with the distributions and the link sizes may take any value with comparable probability. On the other hand, in the high density limit ($Q=10$) the distributions cut off at smaller link sizes and do not depend on the box size:
\begin{equation}
    P_{L_\text{box}}(a) = p'(a), \qquad (Q=10)
\end{equation}
with $p'$ being a suitable function. This is also evident from the analysis of the median $\tilde a_{L_\mathrm{box}}$ of the link size distributions in Fig.~\ref{fig:R_median}(a): for $Q=0$, $\tilde a_{L_\mathrm{box}} \propto L_\mathrm{box}$ and for $Q=10$, $\tilde a_{L_\mathrm{box}} \approx \mathrm{const}$. Consequently, $\tilde a_{L}$, the median computed for the whole simulation cell, characterizes the typical extent of the active region at event onset. This quantity will be referred to as a \emph{dynamic correlation length} $\xi_\mathrm{d} := \tilde a_{L}$.

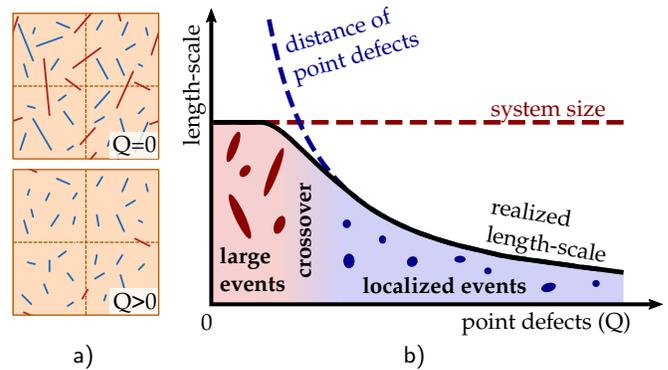
\begin{figure}[!ht]
    \centering
    \includegraphics[width=\columnwidth]{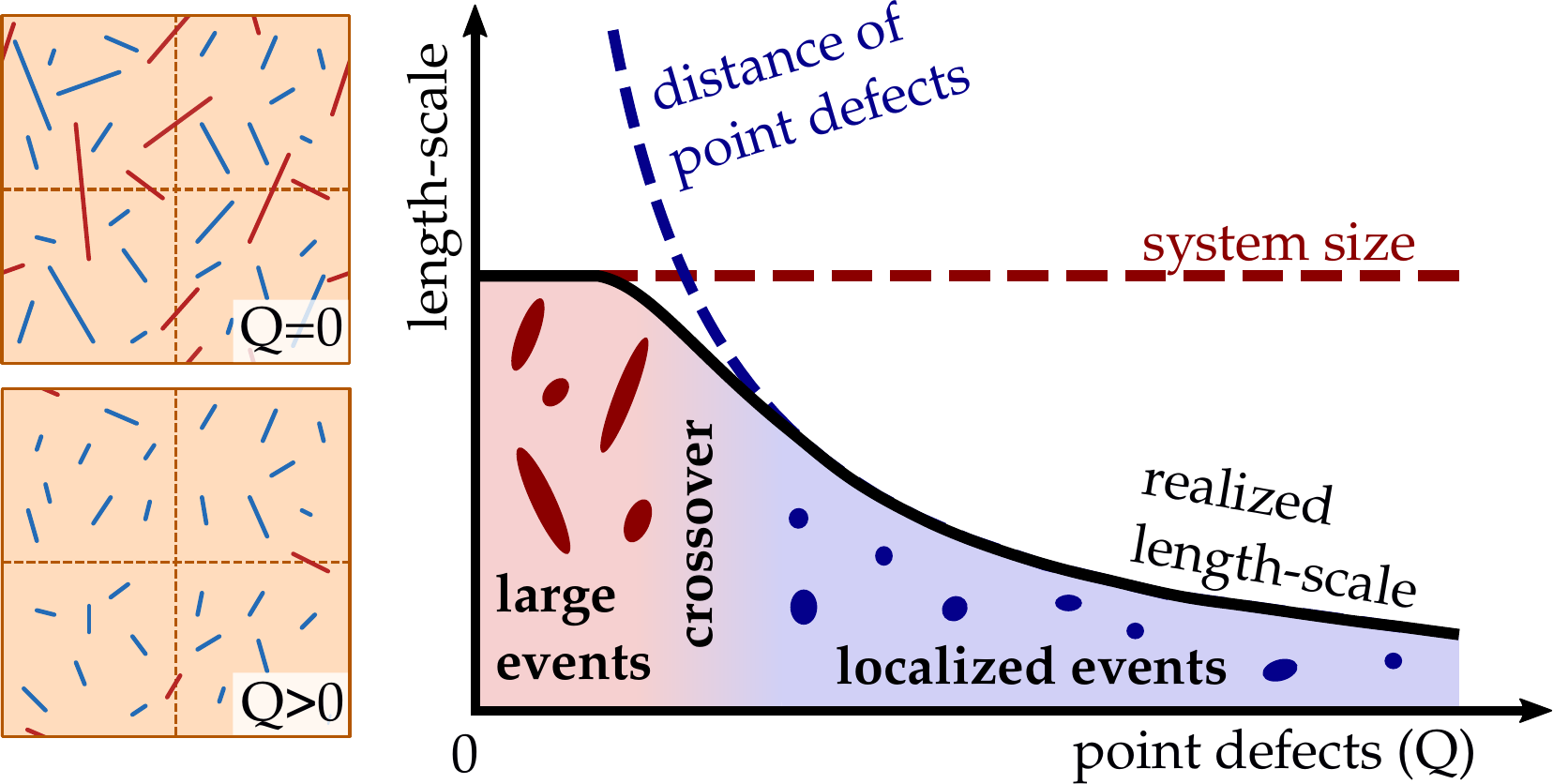}
    \begin{picture}(0,0)
    \put(-100,0){\sffamily{a)}}
    \put(25,0){\sffamily{b)}}
    \end{picture}
    \caption{(a): Schematic representation of link sizes.
    (b): Sketch of the emergent fundamental dynamic length-scale determined by the competition of the length-scales introduced by the system size and short-range interactions (point defects). Where the two lengths are comparable, a smooth crossover describes the realized length-scale.}
    \label{fig:sketch}
\end{figure}

\begin{figure}[!ht]
    \centering
    \includegraphics[width=\columnwidth]{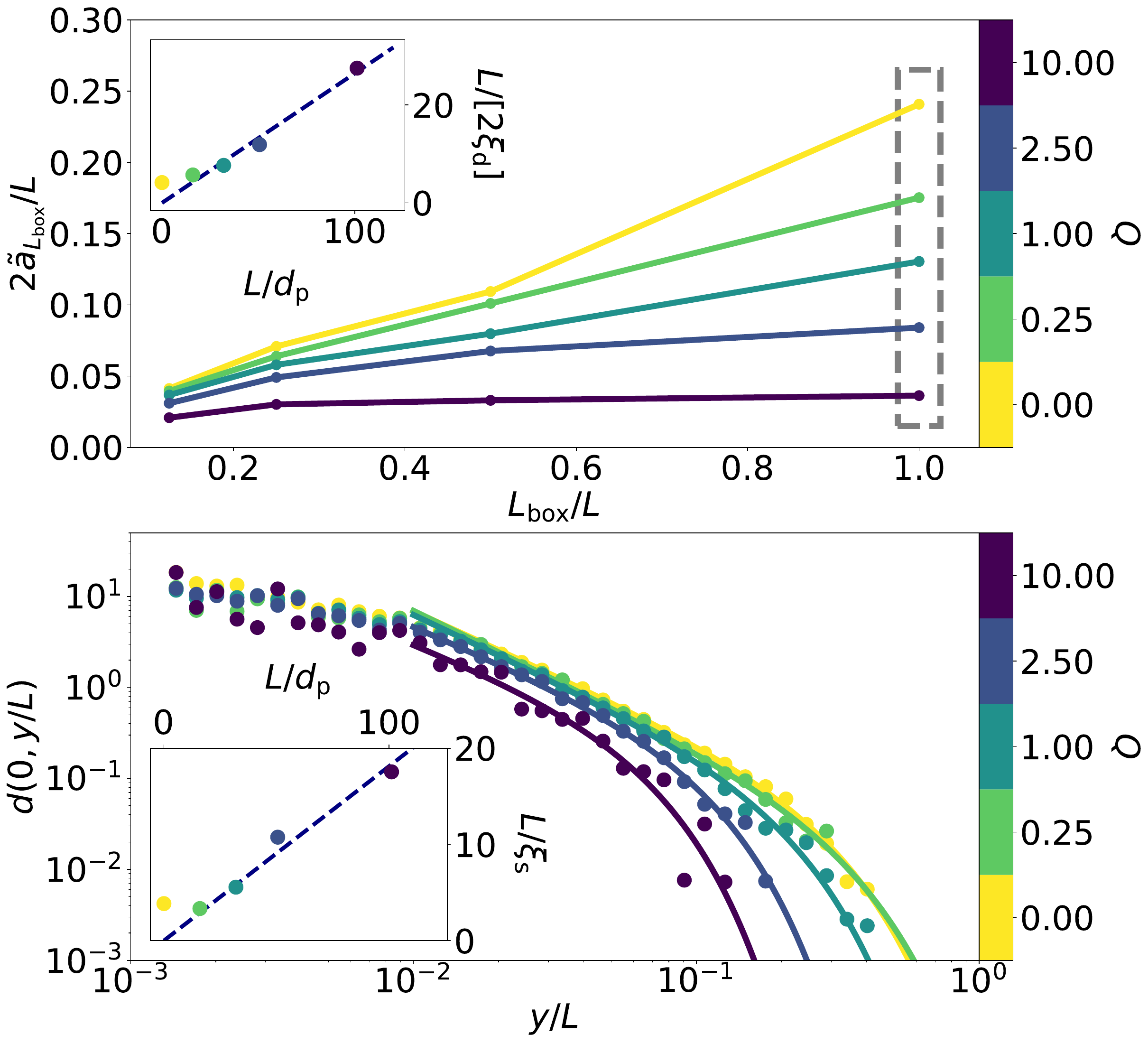}
    \begin{picture}(0,0)
    \put(-125,225){\sffamily{a)}}
    \put(-125,115){\sffamily{b)}}
    \end{picture}
    \caption{(a): Median $\tilde a_{L_\mathrm{box}}$ against the box size $L_\mathrm{box}/L$ for different concentrations $Q$. Note that 
    $\tilde a_{L_\mathrm{box}}$ is proportional with the subbox size $L_\mathrm{box}$ for $Q=0$ and saturates for larger $Q$ values. Inset: the dynamic correlation length $\xi_\mathrm{d}$, that is, the values highlighted in the main panel.
    (b): Two-point correlation $d(0,y/L)$ of +-type dislocations along axis $y$. Inset: the cutoff $\xi_\mathrm{s}$ obtained by fitting as a function of the average point defect distance $d_\mathrm{p}$. 
    For a representative 2D $d(\bm r)$ see Fig.~3 in \cite{supplemental}.
    \label{fig:R_median}
    }
\end{figure}
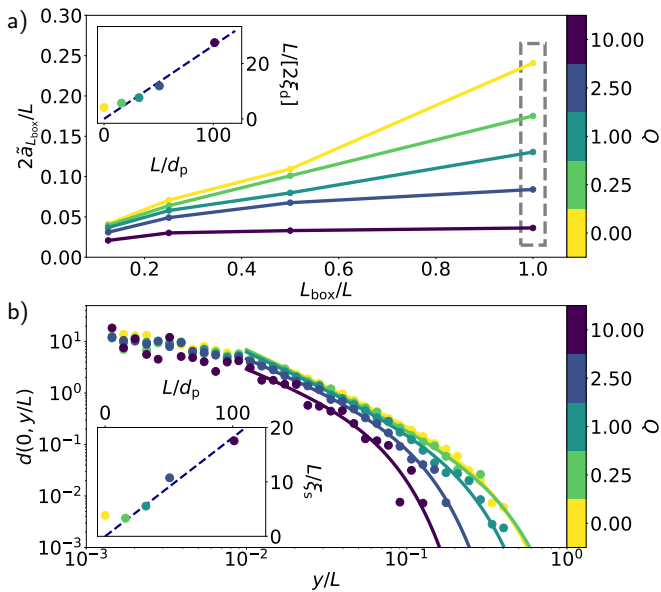

The picture that emerges is as follows. Pure systems are governed by long-range ($\propto 1/r$) interactions and lack natural length-scale. Consequently, avalanches may span the whole system and, as was shown earlier, they have a scale-free size distribution only cut off by the obvious limit posed by the system size \cite{ispanovity2014avalanches}. By adding short-range interactions a natural length-scale is introduced that limits the extension of the avalanches (see the sketch of Fig.~\ref{fig:sketch}(a)). One, thus, concludes that the lower values of the Pearson correlation coefficients in $Q=0$ systems are due to the fact, that if the particular weakest link has large spatial extent, it is likely to get intersected during the subbox division, so, it cannot be activated at the lower level (see transition $B=1 \to 2$ in the $Q=0$ system in Fig.~\ref{fig:box_division}). At high $Q$, however, the link sizes are much smaller, so, such intersections have a much smaller probability, yielding larger correlation values. If fraction $0<f\le 1$ is intersected at a transition $B \to B+1$, then $N_\mathrm{link}(L_\mathrm{box}/2) = (1-f) N_\mathrm{link}(L_\mathrm{box})/4$, together with Eq.~(\ref{eq: dimension def}) yields dimension $D=2-\log_2(1-f)$. Hence, $f=1/2$ (uniform distribution of link sizes) leads to $D=3$, whereas $f=0$ (point-like links) yields $D=2$. These two limits are quite closely realized in pure ($Q=0$) and defect-rich ($Q=10$) systems [Fig.~\ref{fig:round_yield}(d)].

The results indicate that the inclusion of short-range interaction introduces a length-scale to the otherwise scale-free system. A natural candidate for this length-scale is the average spacing of point defects $d_\mathrm{p} = L/\sqrt{N_\mathrm{p}}$. Indeed, according to the inset of Fig.~\ref{fig:R_median}(a) $\xi_\mathrm{d} \propto d_\mathrm{p}$ holds, except for small $Q$ where the typical event size approaches the system size $L$.

It is known that dislocations in pure, equilibrium 2D systems exhibit spatial correlations, that are long-range along axis $y$ \cite{zaiser2001statistical, groma2003spatial, groma2006debye} and have a cut-off if point defects are introduced \cite{ovaska2016collective}. Here we test whether this static correlation length $\xi_\mathrm{s}$ is related to the dynamic correlation length $\xi_\mathrm{d}$. To this end, the two-point correlation functions, defined as $d(\bm r) = \rho_2(\bm r)/\rho^2-1$, are determined from the discrete configurations with $\rho$ and $\rho_2(\bm r)$ being the one- and two-point densities, respectively \cite{zaiser2001statistical, groma2003spatial}. Due to translational invariance, $\rho_2$ only depends on the relative coordinate $\bm r$ of the two dislocations and $\rho=N/L^2$. Figure \ref{fig:R_median}(b) plots these correlation functions along the axis $y$ for different values of $Q$ as well as the fitted functions of the form $d(0,y/L) \propto (y/L)^{-\gamma} \exp(-y/\xi_\mathrm{s})$. The inset yields $\xi_\mathrm{s} \propto d_\mathrm{p}$, that is, $\xi_\mathrm{s} \propto \xi_\mathrm{d}$, so, the static and dynamic correlation lengths are practically identical.

\section{Summary \& outlook}

In this paper we investigated the local yield stress statistics in discrete dislocation systems with and without short-range quenched pinning. The spatial extent of the corresponding plastic events was also analyzed. It was found that the active regions are localized if pinning points are present and can be characterized with a dynamic correlation length $\xi_\mathrm{d}$ being proportional with the average distance of the pinning points. In systems without point defects, however, no such scale exists and plastic events may span the whole system, that is, here $\xi_\mathrm{d} \to L$ (see sketch in Fig.~\ref{fig:sketch}(b)). On a scale above $\xi_\mathrm{d}$, a conventional weakest-link picture is realized: the yield stress of a larger volume is inherited from its weakest subvolume. As such, a cell size equal to the dynamic correlation length can be considered as the RVE. Below $\xi_\mathrm{d}$ (i.e., always in point-defect-free systems), division of the subvolume may lead to the inactivation of the weakest link. However, we emphasize that in pure systems it is not the weakest-link picture that is violated, as also inferred from the obtained Weibull statistics \cite{ispanovity2017role}, rather, the weakest links simply do not have a maximum size. Therefore, the RVE is the whole simulation cell in this case.

From a broader perspective, we first note that yield stress in crystalline materials has always been considered a local quantity. Here we investigated how local it actually is. We found that if only long-range dislocation interactions are present, then yielding is not at all local and yield stress distributions depend on the size of the region the yield stress represents. Short-range effects, however, do introduce an RVE of reduced size that makes yielding indeed local. Similar short-range effects to the one considered here are ubiquitously present in crystals: dislocation reactions, cross-slip, precipitates, solute atoms or various phase or grain boundaries are all expected to introduce a dynamic length scale on an analogous manner. This idea echoes on the long-standing debate on the dominance of either short- \cite{DEVINCRE2001211, MADEC2002689, PhysRevLett.96.125503, sandfeld2015pattern} or long-range interactions \cite{MUGHRABI19831367, groma2016dislocation, ispanovity2020emergence, wu2021cell} in the appearance of dislocation patterns with a characteristic scale. The length-scale $\xi_\mathrm{d}$ may also be related to the concept of ``dislocation mean free path'' introduced in phenomenological plasticity models \cite{devincre2008dislocation}. Here we conclude, that the appearance of the length-scale is, in fact, the result of the competition between long- and short-range effects. The potential in the method of consecutive subbox divisions introduced here is its generalizability to more complex cases to determine the exact value of the dynamic correlation length and, consequently, the size of the RVE. This possibility also applies to other types of heterogeneous materials, such as glasses, and is expected to tackle the issue of RVE selection in the multiscale modeling of complex materials.

\section{Acknowledgments}

\begin{acknowledgments}

Support by the National Research, Development and Innovation Fund of Hungary (contract numbers: NKFIH-FK-138975) is acknowledged. G.P.~was also supported by the ÚNKP-22-4 New National Excellence Program of the Ministry for Culture and Innovation from the source of the National Research, Development and Innovation Fund. D.B.~was also supported by the AMAAL Scholarship of the Association of Hungarian American Academicians, the Hungarian Ministry for
Innovation and Technology and the National Research, Development and Innovation Office.

\end{acknowledgments}

\bibliography{main}

\end{document}